\documentclass[12pt]{article} 
 \RequirePackage[spanish,english]{babel}

\RequirePackage{graphicx,graphics,epsfig}
\begin {document} 
	
\title{Estimaci\'on del n\'umero de reproducci\'on de la epidemia COVID-19 en Culiac\'an Sinaloa, M\'exico}
	
\author{Mart\'{\i}n H. F\'{e}lix-Medina\thanks{%
			mhfelix@uas.edu.mx} \\
		Facultad de Ciencias F\'\i sico-Matem\'aticas\\
		Universidad Aut\'{o}noma de Sinaloa\\
		Ciudad Universitaria, Culiac\'{a}n Sinaloa, M\'{e}xico.}
	
\maketitle

 \selectlanguage{english}
 
\begin{abstract}
Currently the COVID-19 epidemic is developing in the City of Culiac\'an Sinaloa, Mexico, where up to April 20 of 
this year there have been 35 deaths associated with this epidemic. The reproduction number $(R_0)$ of an epidemic 
represents the average number of people infected by an infected person during their period of infection. In this 
work we use the data published by the Secretary of Health of the State of Sinaloa on the number of new infected 
cases confirmed per day and we estimate that the value of $R_0$ is 1.562 with a 95\% confidence interval given by 
(1.401,1.742). We also used the published data on the number of deaths and daily recoveries to estimate the mortality 
rate among the confirmed cases, which turned out to be 16.8\%. We also make a prediction of the number of deaths 
expected to occur on each of the days of the next three weeks from April 20.	
	
- - - - -
	
Actualmente la epidemia COVID-19 se est\'a desarrollando en la Ciudad de Culiac\'an Sinaloa, M\'exico, donde hasta el 
20 de abril del presente a\~no han ocurrido 35 decesos asociados con esta epidemia. El n\'umero de reproducci\'on $(R_0)$ 
de una epidemia representa el n\'umero promedio de personas contagiadas por una persona infectada durante su periodo de 
infecci\'on. En este trabajo usamos los datos publicados por la Secretar{\'\i}a de Salud del Estado de Sinaloa sobre el 
n\'umero de nuevos casos infectados confirmados por d{\'\i}a y estimamos que el valor de $R_0$ es de 1.562 con un intervalo 
del 95\% de confianza dado por (1.401,1.742). Usamos tambi\'en los datos publicados sobre los n\'umeros de decesos y 
recuperados diarios para estimar la tasa de mortalidad entre los casos confirmados, la cual result\'o ser de 16.8\%. 
Realizamos tambi\'en una predicci\'on del n\'umero de decesos que se espera que ocurran en cada uno de los d{\'\i}as de las 
siguientes tres semanas a partir del 20 de abril.
\end{abstract}

 \selectlanguage{spanish}
 
\section{Introducci\'on}

La epidemia COVID-19 inici\'o en Sinaloa el pasado mes de marzo. No obstante la medida ``Qu\'edate en casa'' de 
confinamiento en los hogares adoptada a nivel nacional, la epidemia ha continuado avanzando en Sinaloa y 
particularmente en Culiac\'an, donde se han presentado el mayor n\'umero de casos confirmados y de decesos 
provocados por esta enfermedad. Por tal motivo, es de inter\'es conocer algunos par\'ametros relacionados con 
este brote epid\'emico en esta ciudad. 

Un par\'ametro que permite caracterizar una epidemia es el n\'umero de reproducci\'on o tasa de reproducci\'on 
intr{\'\i}nseca $(R_0)$, el cual representa el n\'umero promedio de personas contagiadas por una persona infectada 
durante su periodo de infecci\'on. Este par\'ametro es importante porque si $R_0>1$, entonces el n\'umero de personas 
contagiadas crecer\'a de manera exponencial, mientras que si $R_0<1$, el n\'umero de contagios disminuir\'a de forma 
exponencial, y consecuentemente, el brote epid\'emico desaparecer\'a r\'apidamente sin producirse muchos contagios.

Estimaciones de $R_0$ correspondientes a epidemias COVID-19 que se presentaron en China y en otros lugares var{\'\i}an entre 
$1.4$ y $6.49$, con media $3.28$ y mediana $2.79$ (v\'ease Liu et al. 2020). Zhang et al. (2020) obtuvieron una estimaci\'on 
de $2.28$ para el caso de la epidemia COVID-19 que se present\'o en el Crucero ``Diamond Princess''. Estos autores utilizaron
el m\'etodo propuesto por Fraser (2007) e implementado en el paquete ``earlyR'' (Jombart et al., 2017) del sistema de c\'omputo 
R (R Core Team, 2020) para obtener esta estimaci\'on. 

En este trabajo, presentamos una estimaci\'on de $R_0$ durante las fases iniciales de la epidemia COVID-19 que est\'a actualmente
desarroll\'andose en la Ciudad de Culiac\'an Sinaloa. Con este prop\'osito utilizamos los datos oficiales de casos confirmados 
publicados por la Secretar{\'\i}a de Salud del Estado de Sinaloa y el paquete ``earlyR'' del sistema de c\'omputo estad{\'\i}stico R.
Presentamos, asimismo, un pron\'ostico del n\'umero de nuevos casos confirmados por d{\'\i}a que se espera que ocurran en las pr\'oximas 
tres semanas. Presentamos tambi\'en una estimaci\'on de la tasa de mortalidad entre los casos confirmados para Culiac\'an, as{\'\i}
como un pron\'ostico del n\'umero de nuevos decesos que se espera que ocurran por d{\'\i}a en las pr\'oximas tres semanas, el cual 
obtuvimos a partir del n\'umero de nuevos decesos y del n\'umero de nuevos casos recuperados ocurridos por d{\'ia} resportados por la
 Secretar{\'\i}a de Salud  Sinaloa.

\section{M\'etodo de Fraser para la estimaci\'on de $R_0$}

El m\'etodo propuesto por Fraser (2007) para estimar $R_0$ en las etapas iniciales de la epidemia consiste en considerar que el 
n\'umero de nuevos infectados $I(t)$ al tiempo $t$ es una variable aleatoria Poisson con media
\begin{equation}
\label{EIt} E\left[I(t)\right]=\int_0^\infty E\left[I(t-\tau)\right]\beta(t,\tau)d\tau,
\end{equation}
donde $\beta(t,\tau)>0$ es la funci\'on de transmisibilidad de la enfermedad por un infectado. Esta funci\'on depende del tiempo 
$t$ que lleva la epidemia y del tiempo $\tau$ al que ocurri\'o el contagio. El n\'umero promedio de personas contagiadas $R_0(t)$ 
al tiempo $t$ por una persona infectada se relaciona con la funci\'on de transmisibilidad $\beta(t,\tau)$ mediante
\begin{equation}
\label{R0t}R_0(t)=\int_0^\infty \beta(t,\tau)d\tau.
\end{equation}
Fraser (2007) supone que 
\begin{equation}
\label{beta}\beta(t,\tau)=\phi_1(t)\phi_2(\tau),
\end{equation}
donde $\phi_2(\tau)>0$ y $\int_0^\infty \phi_2(\tau)d\tau=1$, esto es, que $\phi_2(\tau)$ es una funci\'on de densidad de probabilidad.
Note que este supuesto implica que la transmisibilidad de la enfermedad por un contagiado depende de una funci\'on $\phi_1(t)$ del tiempo $t$
que lleva la epidemia y de la distribuci\'on de probabilidad $\phi_2(\tau)$ del tiempo $\tau$ al que ocurri\'o el contagio. Esta 
funci\'on $\phi_2(\tau)$ se conoce como ``perfil de infectabilidad''. De $(\ref{R0t})$ y $(\ref{beta})$ se sigue que $R_0(t)=\phi_1(t)$, y 
nuevamente de $(\ref{beta})$ se sigue que $\beta(t,\tau)=R_0(t)\phi_2(\tau)$. Por tanto, de $(\ref{EIt})$ se tiene que
$$
E\left[I(t)\right]=R_0(t)\int_0^\infty E\left[I(t-\tau)\right]\phi_2(\tau)d\tau,
$$
y consecuentemente
$$
R_0(t)=\frac{E\left[I(t)\right]}{\int_0^\infty E\left[I(t-\tau)\right]\phi_2(\tau)d\tau}.
$$
Dado que los n\'umeros de nuevos infectados se reportan en tiempos discretos (por ejemplo diariamente), Fraser (2007)  discretiza la funci\'on 
$\phi_2(\tau)$ mediante la funci\'on de probabilidad de masa $w(j)$, $j=1,\ldots,n$, y estima $E\left[I(t)\right]$ por $I(t)$. As{\'\i}, 
estima $R_0(t)$ mediante
$$
\hat R_0(t)=\frac{ I(t) }{ \sum_{j=0}^{n} I(t-j)w(j)}.
$$
Note que para calcular $\hat R_0(t)$ se requiere conocer la funci\'on de probabilidad de masa $w(j)$, $j=1,\ldots,n$. Cori et al. (2013)
indican que en la pr\'actica $w$ se reemplaza por la distribuci\'on de probabilidad del tiempo que transcurre desde que inician los
s{\'\i}ntomas de una persona enferma hasta que inician los s{\'\i}ntomas de una persona contagiada por ella. Esta distribuci\'on se conoce
como la distribuci\'on del intervalo de casos sucesivos (``serial interval'') y se aproxima mediante una distribuci\'on Gamma discretizada 
cuyos par\'ametros se obtienen de casos observados o de reportes en la literatura. Estos autores tambi\'en se\~nalan que este estimador de
$R_0(t)$ es v\'alido a\'un si los valores reportados de $I(t)$ son de nuevos casos confirmados de infectados y no de nuevos casos infectados 
en general (confirmados o no confirmados). Esto es as{\'\i} ya que al calcularse mediante un cociente, si los nuevos casos confirmados son 
una fracci\'on constante del total de nuevos casos infectados, el factor de proporcionalidad se cancela y no afecta el valor del estimador.

Cabe se\~nalar que Cori et al. (2013) usan el modelo de Fraser (2007) para desarrollar un enfoque Bayesiano para estimar $R_0$. Ellos suponen 
una distribuci\'on inicial Gamma para $R_0$ y muestran que la distribuci\'on final de $R_0$ es tambi\'en Gamma. Estos autores indican que las
estimaciones de $R_0(t)$ que se obtienen mediante este enfoque son bastante variables, por lo que se requiere especificar un intervalo de
tiempo (ventana de tiempo) en el que se supone que $R_0(t)$ se mantiene constante. El paquete  ``earlyR'' del sistema R usa la distribuci\'on 
posterior de $R_0$ para construir intervalos de confianza bayesianos de este par\'ametro mediante la simulaci\'on de valores de la 
distribuci\'on posterior.

Nouvellet et al. (2018) usan el modelo Bayesiano de Cori et al. (2013) y desarrollan un procedimiento para pronosticar valores del n\'umero 
de nuevos casos $I(t)$. El procedimiento consiste en generar valores de la distribuci\'on posterior de $I(t)$ mediante la metodolog{\'\i}a de 
simulaci\'on Monte Carlo de Cadenas de Markov (MCMC por sus siglas en Ingl\'es), y a partir de los valores generados calcular los pron\'osticos 
de $I(t)$, as{\'\i} como sus correspondientes intervalos de confianza bayesianos. Este procedimiento fue implementado por Jombart et al. (2018) 
en el paquete ``projections'' del sistema  R.

\section{Estimaci\'on de la tasa de mortalidad entre los casos confirmados}

Aunque idealmente el objetivo es tener informaci\'on sobre la tasa de mortalidad entre la poblaci\'on infectada, este par\'ametro no lo podemos
estimar debido a que desconocemos el n\'umero de infectados. Sin embargo, es factible estimar la tasa de mortalidad entre los casos confirmados
ya que se cuenta con la informaci\'on necesaria. No obstante que este par\'ametro podr{\'\i}a no ser de inter\'es por s{\'\i} mismo, ya que los
casos confirmados es un grupo de personas con caracter{\'\i}sticas muy particulares las cuales dependen del criterio que se use para confirmar
si tienen o no la enfermedad, la estimaci\'on de la tasa de mortalidad de este grupo permite pronosticar el n\'umero de decesos que se espera que 
ocurran en un futuro.

De acuerdo con Ghani et al. (2005), dos estimadores simples de calcular de la tasa de mortalidad al d{\'\i}a $t$ de la epidemia son
$$
\hat p_1(t)=\frac{\sum_{s\le t}D(s)}{\sum_{s\le t}I(s)}
$$
y
\begin{equation}
\label{p2}\hat p_2(t)=\frac{\sum_{s\le t}D(s)}{\sum_{s\le t}R(s)+\sum_{s\le t}D(s)},
\end{equation}
donde $D(s)$, $I(s)$ y $R(s)$ representan los n\'umeros de nuevos decesos, nuevos infectados y nuevos recuperados que ocurren en el d{\'\i}a $s$. 
Note que no necesariamente
$$
\sum_{s\le t}I(s)=\sum_{s\le t}R(s)+\sum_{s\le t}D(s)
$$
ya que al d{\'\i}a $t$ existen personas que est\'an infectadas, pero que no se han recuperado ni han fallecido. Por tanto, el primer estimador $\hat p_1(t)$
no toma en cuenta el estado final (recuperaci\'on o deceso) de esas personas, y consecuentemente, tiende a subestimar la tasa de mortalidad ya que posiblemente
ocurra el deceso de algunas de esas personas. El segundo, estimador $\hat p_2(t)$ corrige el problema de subestimaci\'on al considerar s\'olo los casos para
los cuales se conoce su estado final. Aunque existen otros estimadores de la tasa de mortalidad con mejores propiedades estad{\'\i}sticas que los arriba
se\~nalados, por ejemplo Ghani et al. (2005) proponen un estimador del tipo Kaplan-Meier que toma en cuenta el n\'umero de personas infectadas al d{\'\i}a $t$ 
para las cuales se desconoce su estado final, estos son m\'as dif{\'\i}ciles de calcular, y en estudios num\'ericos se ha encontrado que el estimador $p_2(t)$
tiene un desempe\~no aceptable.

\section{Estimaci\'on de $R_0$ en la etapa inicial de la epidemia COVID-19 en Culic\'an Sinaloa}

El primer caso confirmado de COVID-19 en la Ciudad de Culiac\'an se present\'o el 28 de febrero del presente a\~no. Este caso correpondi\'o 
a una persona originaria del Estado de Hidalgo y que previamente a su llegada a la Ciudad de Culiac\'an estuvo en Italia, donde se presume
que fue contagiado. Desde esa fecha hasta el 19 de abril han transcurrido 52 d{\'\i}as y se han reportado 328 casos confirmados de COVID-19.
La gr\'afica de la Figura 1 muestra el n\'umero de nuevos casos confirmados por d{\'\i}a.

\begin{figure}
	\begin{center}
		\includegraphics[width=5.5in,height=3.0in]{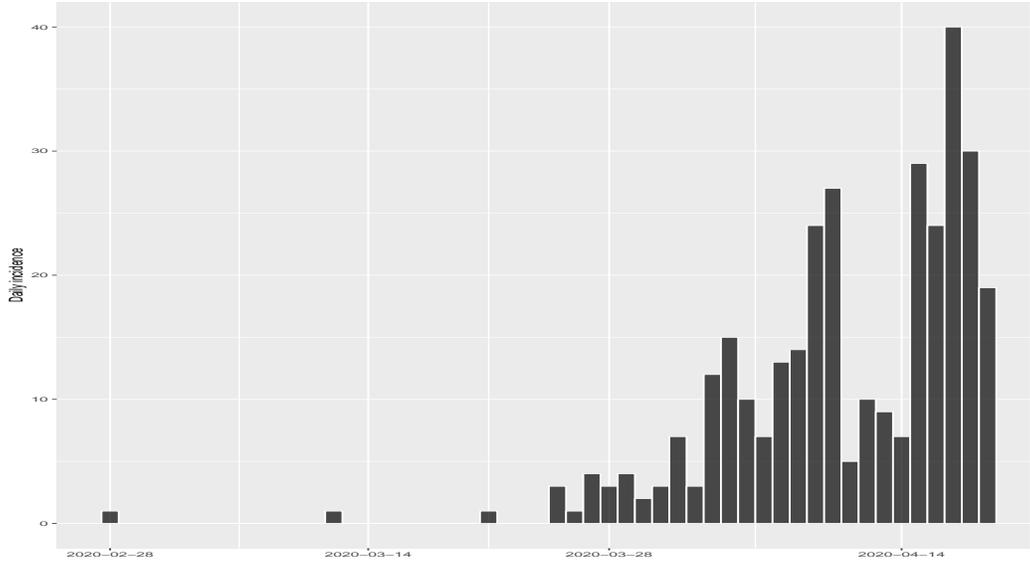}
		\caption{Incidencia de nuevos casos confirmados por d{\'\i}a.}
	\end{center}
\end{figure}

Como se indic\'o en la secci\'on anterior, el m\'etodo de Fraser para estimar $R_0$ requiere que se definan los par\'ametros de la distribuci\'on 
Gamma que describe la distribuci\'on del intervalo de casos sucesivos. Dado que no tenemos informaci\'on que nos permita estimar estos par\'ametros 
para el caso particular de la epidemia que se presenta en Culiac\'an, al igual que en Zhang et al. (2020), usaremos los valores correspondientes
a la epidemia COVID-19 presentada en Wuhan China y reportados por Li et al. (2020). As{\'\i}, consideraremos que la media y la desviaci\'on est\'andar
de esta distribuci\'on son 7.5 y 3.4 d{\'\i}as, respectivamente. Luego, mediante el uso de la funci\'on ``get\_R'' del paquete ``earlyR'', obtenemos
que la estimaci\'on m\'aximo veros{\'\i}mil de $R_0$ es 1.562 (v\'ease la gr\'afica de la Figura 2), esto es, cada persona enferma infecta a alrededor 
de 1.56 personas durante su periodo de contagio. Un intervalo bayesiano del 95\% de confianza de $R_0$, obtenido mediante la simulaci\'on de 10 000 
valores de la distribuci\'on posterior de $R_0$, y generados mediante la funci\'on ``sample\_R'' del paquete ``earlyR'', es $(1.401,1.742)$. As{\'\i}, 
con probabilidad 0.95, el valor de $R_0$ est\'a entre 1.40 y 1.74.

\begin{figure}
	\begin{center}
		\includegraphics[width=5.5in,height=3.0in]{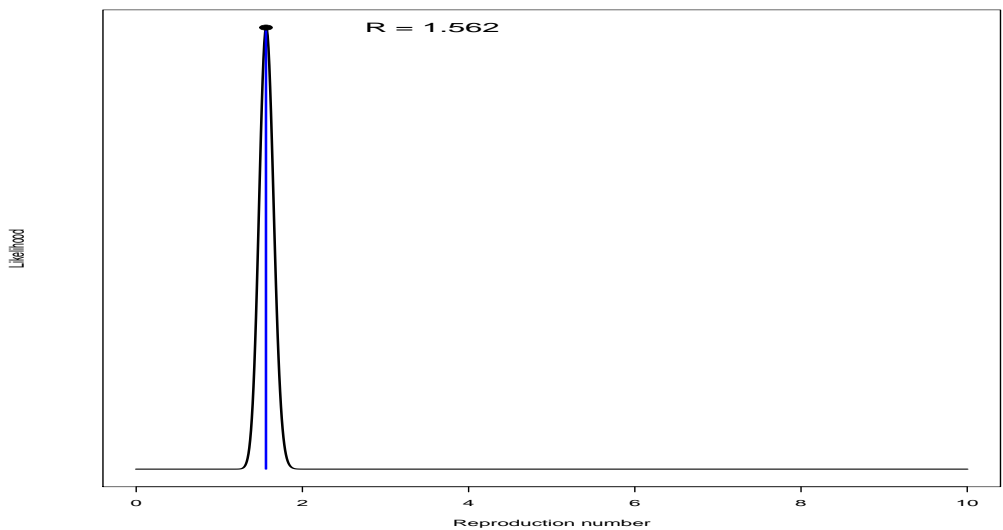}
	\caption{Funci\'on de verosimilitud de $R_0$.}
\end{center}
\end{figure}

Cabe aclarar que la estimaci\'on de $R_0$ se obtuvo bajo el supuesto de que el n\'umero de nuevos casos confirmados por d{\'\i}a es una fracci\'on, que 
se mantiene constante a lo largo del tiempo, del n\'umero total de nuevos casos infectados que ocurren por d{\'\i}a. Si esa fracci\'on no se mantiene
constante, por ejemplo, si al principio de la epidemia el n\'umero de nuevos casos confirmados es una fracci\'on grande del n\'umero de nuevos casos 
infectados, y conforme aumenta el n\'umero de nuevos casos infectados esa fracci\'on disminuye, entonces la estimaci\'on que obtuvimos subestimar\'a el 
verdadero valor de $R_0$. As{\'\i}, es factible que nuestra estimaci\'on sea realmente una cota inferior de $R_0$.

\begin{figure}
	\begin{center}
		\includegraphics[width=5.5in,height=4.0in]{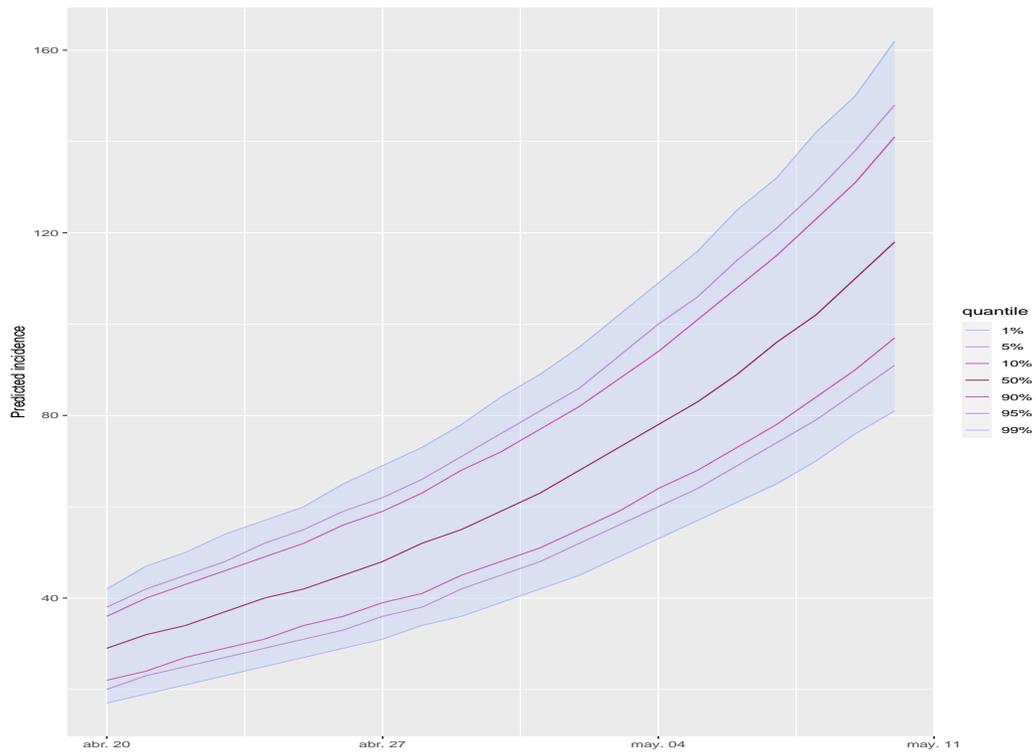}
		\caption{Pron\'osticos de nuevos casos confirmados por d{\'\i}a y bandas bayesianas del 90\%, 95\% y 99\% de confianza.}
	\end{center}
\end{figure}

No obstante que se puede estimar $R_0$ sin conocer el valor exacto del n\'umero de infectados, para poder realizar pron\'osticos del n\'umero de
nuevos infectados por d{\'\i}a mediante el m\'etodo propuesto por Nouvellet et al. (2018) se requiere conocer el n\'umero de nuevos infectados $I(t)$ 
por d{\'\i}a. Sin embargo, es posible pronosticar el n\'umero de nuevos casos confirmados por d{\'\i}a que se presentar\'an en el futuro. Este ejercicio 
es importante porque muestra la tendencia que se espera que tengan los nuevos casos confirmados y, bajo el supuesto de que los nuevos casos confirmados 
son una fracci\'on constante de los nuevos casos infectados, se sigue que los nuevos casos infectados tendr\'an una tendencia similar. En la Figura 3 se 
muestra la gr\'afica de los valores pronosticados de los nuevos casos confirmados, calculados mediante la mediana de la distribuci\'on posterior de $I(t)$, 
(curva correspondiente al cuantil 50\%), as{\'\i} como las bandas bayesianas de confianza del 90\%, 95\% y 99\%, obtenidas mediante la funci\'on ``project'' 
del paquete``projections'' y usando 10 000 valores simulados de la distribuci\'on posterior para cada d{\'\i}a.

Como se observa en la gr\'afica de la Figura 3, la tendencia del n\'umero de nuevos casos confirmados, y por tanto, la de los nuevos casos infectados, es 
creciente hasta mayo 10. Por lo que, de acuerdo con este modelo, si se mantienen las condiciones actuales, el n\'umero de nuevos infectados por d{\'\i}a 
continuar\'a creciendo de forma exponencial.

\section{Estimaci\'on de la tasa de mortalidad entre los casos confirmados de la epidemia COVID-19 en Culic\'an Sinaloa}

En la gr\'afica de la Figura 4 se muestra la incidencia de nuevos decesos diarios registrados hasta abril 19. Las estimaciones de la tasa de mortalidad 
diarias entre los casos confirmados, calculadas mediante el estimador $(\ref{p2})$ se muestran en la gr\'afica de la Figura 5. Dado que al d{\'\i}a 19 
de abril se ten{\'\i}an 173 casos recuperados y 35 decesos, la estimaci\'on de la tasa de mortalidad para ese d{\'\i}a es de 35/(173+35)=0.168. Observe 
que la alta variabilidad de los valores de las estimaciones de la tasa de mortalidad diaria se debe a que al principio los totales de los casos recuperados 
y de decesos son relativamente peque\~nos, y conforme estos van aumentando, la estimaci\'on de la tasa de mortalidad se va aproximando a su valor real.

\begin{figure}
	\begin{center}
		\includegraphics[width=5.5in,height=3.0in]{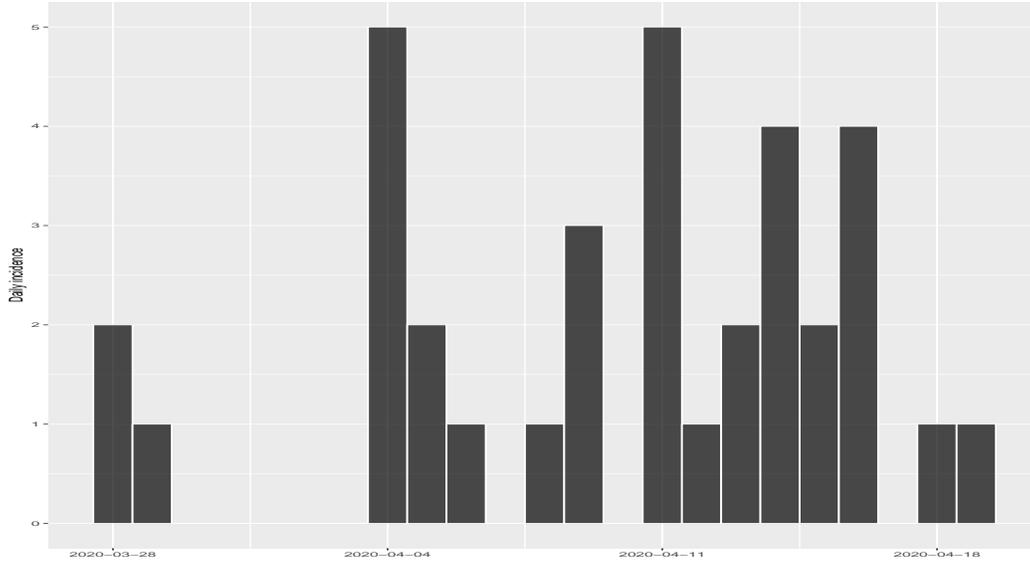}
		\caption{Incidencia de nuevos decesos diarios.}
	\end{center}
\end{figure}

\begin{figure}
	\begin{center}
		\includegraphics[width=5.5in,height=3.0in]{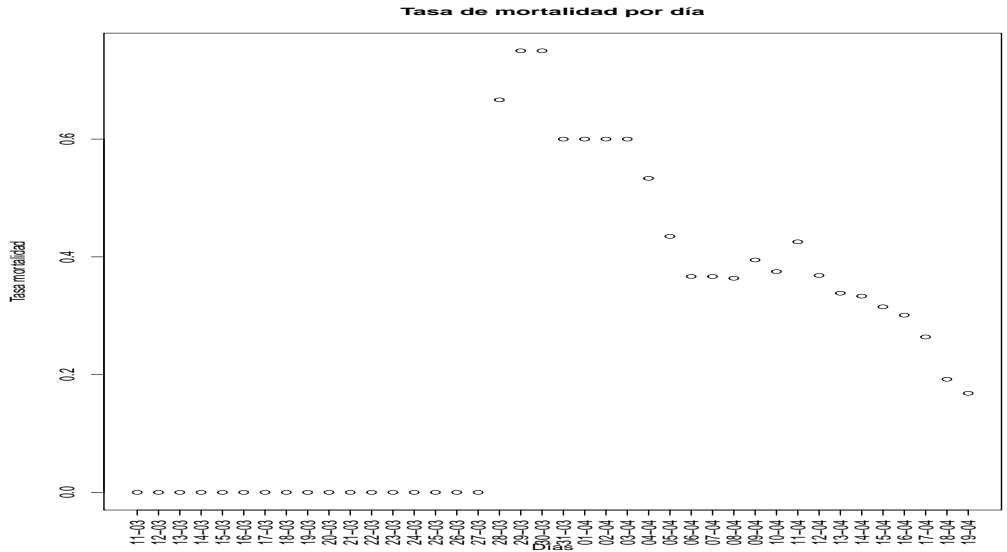}
		\caption{Tasa de mortalidad diaria entre casos confirmados.}
	\end{center}
\end{figure}

Si consideramos que la tasa de mortalidad se mantiene constante en 0.168, podemos pronosticar los nuevos decesos por d{\'\i}a que ocurrir\'an en las 
pr\'oximas tres semanas multiplicando los valores pronosticados de los nuevos casos confirmados por 0.168. Los valores pronosticados y las correspondientes 
bandas bayesianas del 95\% de confianza se muestran en la gr\'afica de la Figura 6. Obs\'ervese que para el 10 de mayo estar{\'\i}an ocurriendo entre 15 y 
25 nuevos decesos diarios con probabilidad 0.95.

\begin{figure}
	\begin{center}
		\includegraphics[width=5.5in,height=4.0in]{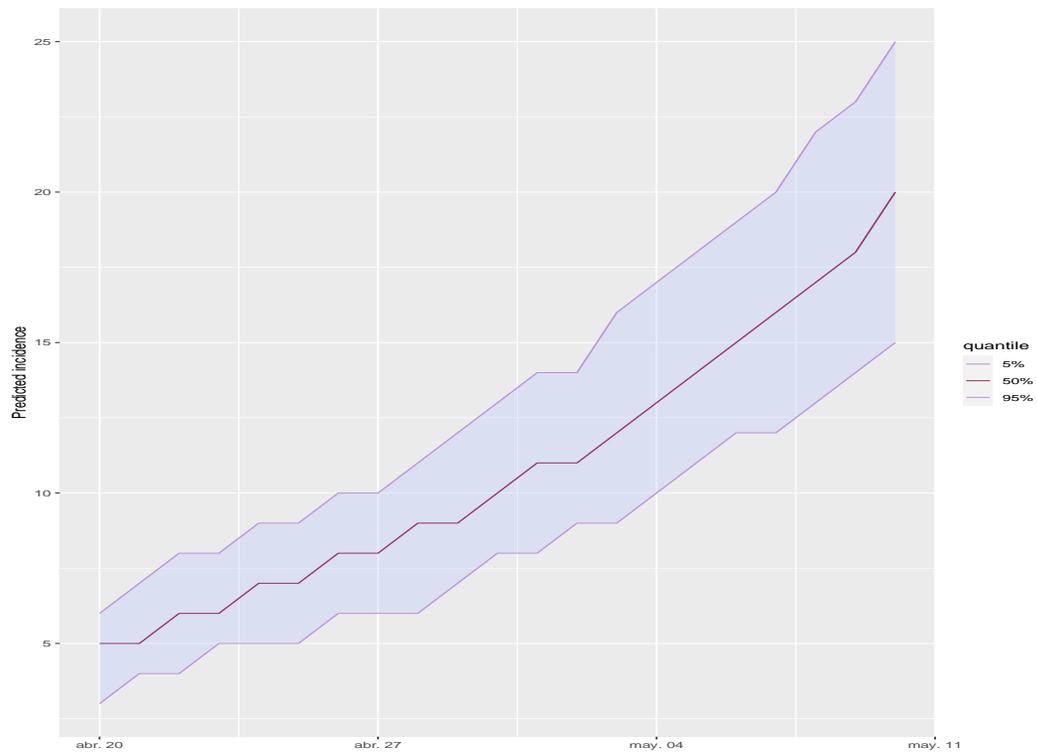}
		\caption{Pron\'osticos de nuevos decesos por d{\'\i}a y bandas bayesianas del 95\% de confianza.}
	\end{center}
\end{figure}

En la gr\'afica de la Figura 7 se muestran los pron\'osticos del n\'umero de decesos acumulados para cada d{\'i}a y las correspondientes bandas bayesianas 
del 95\% de confianza que se espera que ocurran durante las pr\'oximas tres semanas.

\begin{figure}
	\begin{center}
		\includegraphics[width=5.5in,height=4.0in]{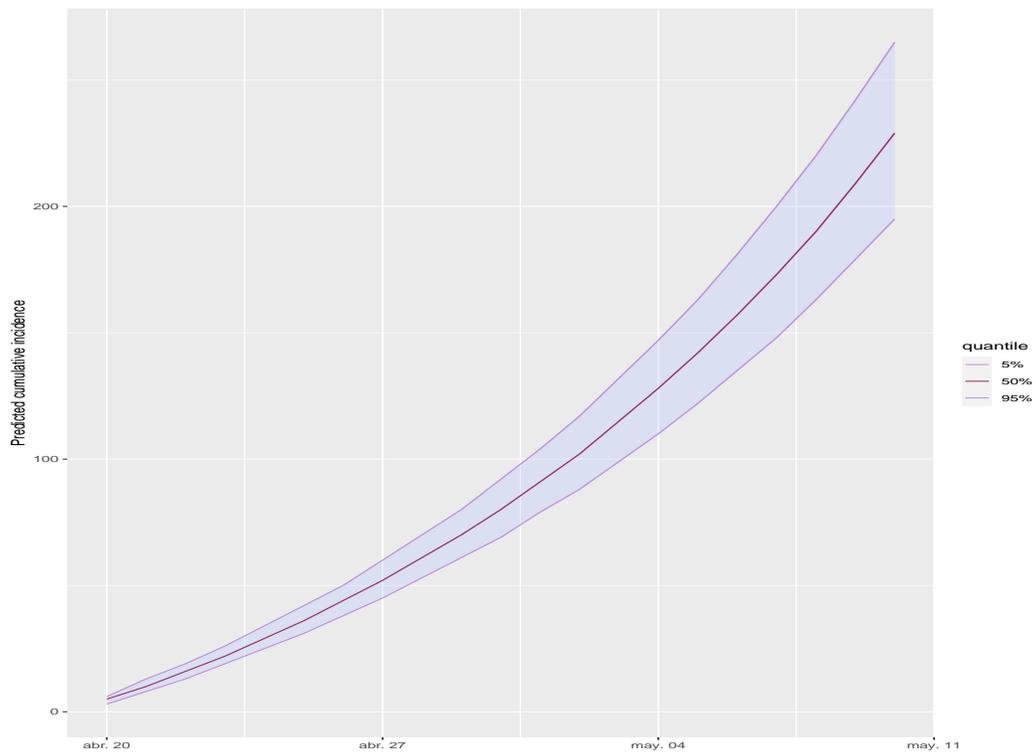}
		\caption{Pron\'osticos de decesos acumulados para cada d{\'\i}a y bandas bayesianas del 95\% de confianza.}
	\end{center}
\end{figure}

\section{Conclusiones}

En este trabajo se usaron los datos publicados por la Secretar{\'\i}a de Salud del Estado de Sinaloa sobre el n\'umero de nuevos casos confirmados diarios de 
COVID-19 en Culiac\'an Sinaloa para estimar el n\'umero promedio de personas que contagia una persona infectada durante su periodo de contagio $(R_0)$. La 
estimaci\'on que se obtuvo fue de 1.56. Esta estimaci\'on se obtuvo bajo el supuesto de que el n\'umero de nuevos casos confirmados diarios es una fracci\'on 
constante del n\'umero de nuevos contagios diarios que ocurren. Si esa fracci\'on no es constante, sino que conforme aumenta el n\'umero de nuevos casos
infectados diarios la fracci\'on de nuevos casos confirmados disminuye, lo cual es factible que ocurra, la estimaci\'on que se obtuvo es una subestimaci\'on 
del verdadero valor de $R_0$. 

Se considero tambi\'en la estimaci\'on de la tasa de mortalidad diaria entre los casos confirmados usando los n\'umeros de nuevos decesos diarios reportados 
y de nuevos casos recuperados diarios. Estos \'ultimos n\'umeros se obtuvieron substrayendo del total de casos infectados acumulados para cada d{\'\i}a, la 
suma del n\'umeros de casos activos (infectados confirmados) que se tienen por d{\'\i}a (los cuales son reportados por la Secretar{\'\i}a de Salud) y el 
n\'umero de decesos acumulados para cada d{\'ia}. La estimaci\'on que se obtuvo para el d{\'\i}a 19 de abril fue de 0.168. Este valor puede parecer alto si 
se compara con las que se han publicado para otros lugares. Por ejemplo, Baud et al. (2020) estiman que el 1 de marzo la tasa de mortalidad en China fue de 
0.056 con un intervalo del 95\% de confianza de $(0.54,0.58)$, mientras que para el resto de los pa{\'\i}ses (tomados en conjunto) fue de 0.152 con un 
intervalo del 95\% de confianza de $(0.125,0.179)$. Existen dos razones por las cuales el valor que obtuvimos para Culiac\'an sea realtivamente grande. 
Una es que la estimaci\'on corresponde a la tasa de mortalidad entre los casos confirmados y no entre los casos infectados. Es factible que una buena parte 
de los casos confirmados est\'e formada por personas que requieren atenci\'on hospitalaria, y por tanto, tengan un mayor riesgo de muerte. La otra es que 
los datos sobre el n\'umero de nuevos casos recuperados por d{\'\i}a no sean actualizados diariamente por la Secretar{\'\i}a de Salud, y por tanto, el 
denominador de la expresi\'on de $\hat p_2(t)$ sea menor que el valor real. En este caso, nuestra estimaci\'on ser{\'\i}a una sobreestimaci\'on.

Vale la pena se\~nalar que para obtener una estimaci\'on de la tasa de mortalidad entre todos los infectados se requiere estimar el total de infectados en
la poblaci\'on. Una manera de hacerlo ser{\'\i}a aplicar el cuestionario que dise\~n\'o la Secretar{\'\i}a de Salud de Sinaloa para identificar un caso 
sospechoso a una muestra de familias seleccionadas mediante marcaci\'on aleatoria de d{\'\i}gitos telef\'onicos. Con esta muestra se podr{\'\i}a estimar la 
fracci\'on de casos sospechos en la poblaci\'on y con informaci\'on sobre la proporci\'on de casos confirmados que se obtienen de los casos sospechosos se 
podr{\'\i}a estimar la fracci\'on de casos infectados en la poblaci\'on. Con esta estimaci\'on y con informaci\'on sobre el tama\~no de la poblaci\'on en
Culiac\'an se podr{\'\i}a estimar el total de casos infectados en el municipio.

\section*{Referencias}

\begin{description}
	
\item[] Baud, D., Qi, X., Nielsen-Saines, K., Musso, D., Pomar, L., and Favre, G. (2020). Real estimates of mortality following COVID-19 infection. 
{\it The Lancet Infectious Diseases}. doi:https://doi.org/10.1016/S1473-3099(20)30195-X.
	
\item[] Cori, A., Ferguson, N. M., Fraser, C., Cauchemez, S. (2013). A new framework and software to estimate time-varying reproduction numbers during 
epidemics. {\it American Journal of Epidemiology}, {\bf 178}, 1505–1512.	

\item[] Fraser, C. (2007). Estimating individual and household reproduction numbers in an emerging epidemic. {\it PLoS One}, 2(1):e758.
doi:10.1371/journal.pone.0000758.

\item[] Ghani, A. C., Donnelly, C. A., Cox, D. R., Griffin, J. T., Fraser, C., Lam, T. H., Ho, L. M., Chan, W. S., Anderson, R. M., Hedley, A. J., and
Leung, G. M. (2005). Methods for estimating the case fatality ratio for a novel, emerging infectious disease. {\it American Journal of  Epidemiology},
{\bf 162}, 479-486

\item[] Jombart, T., Cori, A., and Nouvellet, P. (2017). earlyR: Estimation of transmissibility in the early stages of a disease outbreak. Available 
from: https://CRAN.R-project.org/package =earlyR.

\item[] Jombart, T., Nouvellet, P., Bhatia, S., and Kamvar, Z. N. (2018). projections: Project future case incidence. Available from: 
https://CRAN.R-project.org/package=projections.

\item[] Li Q, Guan X., Wu P., et al. (2020). Early transmission dynamics in Wuhan, China, of novel coronavirus - infected pneumonia. {\it The New England
	Journal of Medicine}, {\bf 382}, 1199-1207. doi: 10.1056/NEJMoa2001316.
	
\item[] Liu, Y., Gayle, A. A., Wilder-Smith, A., and Rocklov, J. (2020). The reproductive number of COVID-19 is higher compared to SARS coronavirus.
{\it Journal of Travel Medicine}. doi: 10.1093/jtm/taaa021.
	
\item[] Nouvellet, P., Cori, A., Garsk,e T., Blake, I. M., Dorigatti, I., Hinsley, W., Jombart, T., Mills, H. L., Nedjati-Gilani, G., Van Kerkhove, M. D.,
Fraser, C., Donnelly, C. A., Ferguson, N. M., and Riley, S. (2018). A simple approach to measure transmissibility and forecast incidence. {\it Epidemics},
{\bf 22}, 29–35.

\item[] R Core Team (2020). R: A language and environment for statistical computing. R Foundation for Statistical Computing, Vienna, Austria. 
URL https://www.R-project.org/.
	
\item[] Zhang, S., Diao, M., Yu, W., Pei, L., Lin, Z., and Chen, D. (2020). Estimation of the reproductive number of novel coronavirus (COVID-19) and 
the probable outbreak size on the Diamond Princess cruise ship: A data-driven analysis. {\it International Journal of Infectious Diseases},{\bf 93},
201-204.
	
\end{description}

\end{document}